\documentclass[a4paper,12pt]{article}
\usepackage{amssymb}
\usepackage{amsmath}
\usepackage{amsthm}
\usepackage{latexsym}
\begin{document}
\title{\Large\textbf{Time-Dependent Automorphism Inducing
Diffeomorphisms, Open Algebras and the Generality of the
Kantowski-Sachs Vacuum Geometry}}
\author{\textbf{T. Christodoulakis}
\thanks{e-mail:tchris@cc.uoa.gr}~~\textbf{\&~G.O. Papadopoulos}
\thanks{e-mail: gpapado@cc.uoa.gr}}
\date{}
\maketitle
\begin{center}
\textit{University of Athens, Physics Department\\
Nuclear \& Particle Physics Section\\
Panepistimioupolis, Ilisia GR 157--71, Athens, Hellas}
\end{center}
\vspace{1cm}
\numberwithin{equation}{section}
\begin{abstract}
Following the spirit of a previous work of ours, we
investigate the group of those General Coordinate
Transformations (GCTs) which preserve manifest spatial homogeneity.
In contrast to the case of Bianchi Type Models we, here, permit an
isometry group of motions $G_{4}=SO(3)\otimes T_{r}$, where $T_{r}$ is
the translations group, along the radial direction, while $SO(3)$ acts
multiply transitively on each hypersurface of simultaneity $\Sigma_{t}$.
The basis 1-forms, can not be invariant under the action of
the entire isometry group and hence produce an Open Lie Algebra. In order for
these GCTs to exist and have a non trivial, well defined action, certain
integrability conditions have to be satisfied; their solutions, exhibiting
the maximum expected ``gauge'' freedom, can be used to simplify the generic,
spatially homogeneous, line element. In this way an alternative proof of the
generality of the Kantowski-Sachs (KS) vacuum is given, while its most
general, manifestly homogeneous, form is explicitly
presented.
\end{abstract}
\newpage
\section{Introduction}
In a previous work \cite{ChrisJMP} we have found and studied the
group of those General Coordinate
Transformations (GCTs) which leave the line element of a generic Bianchi
Type Geometry, quasi-form invariant; i.e.
preserve manifest spatial homogeneity. It was found that these GCTs, which
mix time and space variables in the new space
variables, induce special time-dependent automorphic changes on the spatial
scale factor matrix $\gamma_{\alpha\beta}(t)$,
along with the corresponding changes on the lapse function $N(t)$ and the
shift vector $N^{\alpha}(t)$.
These transformations --called Time-Dependent Automorphism Inducing
Diffeomorphisms (AIDs)-- contain the maximum expected ``gauge'' freedom i.e.
4 arbitrary functions of time for each and every Bianchi Type. Using this
freedom one can significantly simplify the generic line element and thus the
resultant Einstein's Field Equations (EFEs). Also, the number of
the expected essential constants calculated in each  case, is in agreement
with the suggested, corresponding, number in the
established literature (e.g. \cite{Wainwright,Kramer} and the references
therein).

A basic point of this analysis, is that Bianchi Type Geometries, are
characterized by the existence of a 3-dimensional
isometry group of motions $G_{3}$, which acts simply transitively on each
hypersurface of simultaneity $\Sigma_{t}$.
This means that there exists an invariant basis of 1-forms
$\sigma^{\alpha}_{i}(x)$, satisfying:
\begin{equation} \label{rotation}
\sigma^{\alpha}_{i,j}(x)-\sigma^{\alpha}_{j,i}(x)=
2~C^{\alpha}_{\mu\nu}\sigma^{\mu}_{j}(x)\sigma^{\nu}_{i}(x)
\end{equation}
where $C^{\alpha}_{\mu\nu}$ are the --space independent-- structure
constants of the corresponding isometry group
$G_{3}$.

In the present letter we wish to extent this analysis in the case of open
algebras; i.e. when the $C^{\alpha}_{\mu\nu}$s, are space dependent and thus
become structure functions. This usually happens when the isometry
group of motions $G_{r}$, acts multiply transitively on each hypersurface of
simultaneity $\Sigma_{t}$ --see \cite{Wainwright, Ryan} for more details on
the subject.

In 3 dimensions the KS Type spaces are defined \cite{Sachs, Kantowski} as
those admitting an isometry group $G_{4}$ which acts on spacelike
hypersurfaces, with no subgroup $G_{3}$ acting simply transitively on the
hypersurfaces --but instead, an Abelian subgroup $G_{2}$, acting on these.
Because each four dimensional Lie algebra contains a three dimensional
subalgebra \cite{Kantowski}, there exists a three dimensional isometry group
if a four dimensional one exists. Thus also for the KS model, there is a
$G_{3}$ subgroup which however acts simply transitively only on two
dimensional spacelike surfaces which consequently are of constant curvature
(maximally symmetric two dimensional spaces). Kantowski \cite{Kantowski}
showed that two dimensional surfaces of zero and negative
curvature give rise to four dimensional invariance groups, which have simply
transitive three dimensional subgroups. Thus the only
possibility left is that of the two dimensional surfaces of positive
constant curvature, i.e. two dimensional spheres.

The Killing Vector Fields (KVFs), in $r,\theta,\phi$ coordinates, are:
\begin{equation} \label{Killing}
\begin{array}{ll}
K_{1}=-Cos(\phi)\frac{\partial}{\partial\theta}+Cot(\theta)
Sin(\phi)\frac{\partial}{\partial\phi}
&
K_{2}=Sin(\phi)\frac{\partial}{\partial\theta}+Cot(\theta)
Cos(\phi)\frac{\partial}{\partial\phi}\\
K_{3}=\frac{\partial}{\partial\phi} & K_{4}=\frac{\partial}{\partial r}
\end{array}
\end{equation}
The corresponding Lie algebra, is:
\begin{equation} \label{LieAlgebra}
\begin{array}{cc}
\{K_{i},K_{j}\}=\varepsilon^{k}_{ij}K_{k}, ~~~ i,j,k, \in \{1,2,3\}\\
&\\
\{K_{4},K_{j}\}=0,  ~~~j \in \{1,2,3\}
\end{array}
\end{equation}
where $\varepsilon^{k}_{ij}$ is the Levi-Civita symbol. One may
see that, $K_{1}, K_{2}, K_{3}$ correspond to the $G_{3}$ which
acts on two dimensional spacelike surfaces of positive constant
curvature (two spheres), while $K_{3}, K_{4}$ correspond to the
Abelian $G_{2}$ which acts on the three dimensional spacelike
hypersurfaces. If one demands that the generic 3-metric
$g_{ij}(x)$ be invariant under all KVFs i.e.
$\pounds_{K}g_{ij}=0$, one obtains the form
$g_{ij}(x)=A\delta_{\alpha\beta}
\sigma^{\alpha}_{i}(x)\sigma^{\beta}_{j}(x)$ where $A$ is a
positive number, $\delta_{\alpha\beta}$ is the identity 3-matrix
and $\sigma^{\alpha}_{i}(x)$ a basis of 1-forms, which is
invariant under the action of the $G_{2}$ only:
\begin{equation} \label{oneforms}
\sigma^{\alpha}_{i}(x)=\left(\begin{array}{ccc}
1 & 0 &0\\
0 & 1 & 0\\
0 & 0 & Sin(\theta)
\end{array}\right)
\end{equation}
They satisfy:
\begin{equation} \label{DeRham}
d\sigma^{\alpha}=C^{\alpha}_{\mu\nu}\sigma^{\mu}
\wedge\sigma^{\nu}\Leftrightarrow
\sigma^{\alpha}_{i,j}(x)-\sigma^{\alpha}_{j,i}(x)=
2~C^{\alpha}_{\mu\nu}(x)\sigma^{\mu}_{j}(x)\sigma^{\nu}_{i}(x)
\end{equation}
where $C^{3}_{23}(x)=Cot(\theta)/2$ is the only non vanishing, structure
function.

If one considers transformations of the form
$x^{i}=f^{i}(\widetilde{x}^{j})$ and demands the manifest
splitting to be preserved, then when an invariant basis exists,
one arrives at a 3-metric
$\widetilde{g}_{ij}(\widetilde{x})=\widetilde{\gamma}_{\mu\nu}
\sigma^{\mu}_{i}(\widetilde{x})\sigma^{\nu}_{j}(\widetilde{x})$
--see \cite{ChrisCMP} for details--, where
$\widetilde{\gamma}_{\mu\nu}=
\Lambda^{\alpha}_{\mu}\Lambda^{\beta}_{\nu}A\delta_{\alpha\beta}$
and $\Lambda^{\alpha}_{\mu}$ must belong to the automorphism group
of the Lie Algebra i.e. must satisfy $C^{\xi}_{\alpha\beta}
\Lambda^{\alpha}_{\mu}\Lambda^{\beta}_{\nu}
=C^{\omega}_{\mu\nu}\Lambda^{\xi}_{\omega}$. For the case of the
basis (\ref{oneforms}), the elements of which form an open
algebra, the notion of automorphisms can be extended; they must be
the matrices satisfying (\ref{systemlambda})with zero left-hand
side. Thus the following admissible matrices
$\Lambda^{\alpha}_{\beta}$s are obtained:
\begin{equation}
\Lambda^{\alpha}_{\mu}=\left(\begin{array}{ccc}
\lambda_{1} & \lambda_{2} & 0\\
0 & \lambda_{5} & 0\\
0 & \lambda_{8} & \lambda_{9}\end{array}\right)
\end{equation}
with the condition
$x^{2}=ArcCot(Cot(\widetilde{x}^{2})/\lambda_{5})$. Notice that
the subgroup of the above transformations designated by
$\lambda_{5}=1$, is exactly what one would call automorphisms from
the cotangent space point of view i.e. elements of $GL(3, \Re)$
which leave the structure function tensor invariant in form and
value. The most general 3-metric is thus given by the basis
(\ref{oneforms}) and the matrix:
\begin{equation} \label{generalhomogeneous}
\widetilde{\gamma}_{\mu\nu}=\Lambda^{\alpha}_{\mu}
\Lambda^{\beta}_{\nu} A\delta_{\alpha\beta}=
\left(\begin{array}{ccc}
\widetilde{\gamma}_{11} & \widetilde{\gamma}_{12} & 0\\
\widetilde{\gamma}_{12} & \widetilde{\gamma}_{22} & \widetilde{\gamma}_{23} \\
0 & \widetilde{\gamma}_{23} & \widetilde{\gamma}_{33}\end{array}\right)
\end{equation}
It is note worthy that the $(1,3)$ component of the matrix
(\ref{generalhomogeneous}), can not acquire a non zero value
through the diffeomorphisms considered. One might think that
other, more general coordinate transformations might fill this gap
but this is not true: if one calculates the Ricci scalar of a
3-metric with such a non zero component of
$\widetilde{\gamma}_{\alpha\beta}$ --in the given basis one
forms--, one would find it to depend on $\widetilde{x}^{2}$ and
therefore, would conclude that this metric is not spatially
homogeneous.

\section{AIDs and the generality of the KS vacuum model}
In the 3+1 decomposition --see e.g. \cite{Misner}-- one makes use of the
lapse function $N$ and the shift vector $N^{i}$ and sees the 4-dimensional
line element parameterized as:
\begin{equation} \label{lineelement}
ds^{2}=(N^{i} N_{i}-N^{2})dt^{2}+2N_{i}dx^{i}dt+g_{ij}dx^{i}dx^{j}
\end{equation}
EFEs corresponding to (\ref{lineelement}), expressed in terms
of the extrinsic curvature:
\begin{subequations} \label{Einsteinequations}
\begin{displaymath}
K_{ij}=\frac{1}{2N}(N_{i\mid j}+N_{i\mid j}-\frac{\partial
g_{ij}}{\partial t})
\end{displaymath}
\textrm{are:}
\begin{equation} \label{G00}
H_{0}=K^{i}_{j}K^{j}_{i}-K^{2}+R=0
\end{equation}
\begin{equation}\label{G0i}
H_{i}=K^{j}_{i\mid j}-K_{\mid i}=0
\end{equation}
\begin{equation} \label{Gij}
\partial_{t}K^{i}_{j}-NKK^{i}_{j}+NR^{i}_{j}+g^{il}N_{\mid jl}-
(K^{i}_{j\mid l}+K^{i}_{l}N^{l}_{\mid j}-
K^{l}_{j}N^{i}_{\mid l})=0
\end{equation}
\end{subequations}

If one restricts attention to spatially homogeneous spacetimes,
and uses a set of basis 1-forms to decompose the spatial part of
the metric and the shift vector, one will find that the line
element (\ref{lineelement}) assumes the form:
\begin{equation} \label{lineelementoneform}
ds^{2}=(N^{\alpha}(t) N_{\alpha}(t)-N^{2}(t))dt^{2}+
2N_{\alpha}(t)\sigma^{\alpha}_{i}(x)dtdx^{i}
+\gamma_{\alpha\beta}(t)\sigma^{\alpha}_{i}(x)
\sigma^{\beta}_{j}(x)dx^{i}dx^{j}
\end{equation}
Latin indices are spatial, with domain of definition $\{1,2,3\}$, while
Greek indices number the different basis 1-forms,
with the same domain of definition, and are lowered and raised
by $\gamma_{\alpha\beta}(t)$ and
$\gamma^{\alpha\beta}(t)$, respectively.

Insertion of relations (\ref{DeRham}), (\ref{lineelementoneform}),
into (\ref{Einsteinequations}) results in the following set of
Ordinary Differential Equations (ODEs):
\begin{subequations} \label{SpatialEinstein}
\begin{equation} \label{Quadratic}
E_{0}\doteq K^{\alpha}_{\beta}K^{\beta}_{\alpha}-K^{2}+R=0
\end{equation}
\begin{equation} \label{Linear}
E_{i}\doteq \left(2K^{\epsilon}_{\beta}C^{\rho}_{\epsilon\rho }-2
K^{\epsilon}_{\rho}C^{\rho}_{\beta\epsilon}+
K^{\alpha}_{\beta,j}\sigma^{j}_{\alpha}\right)\sigma^{\beta}_{i}(x)-K_{,i}=0
\end{equation}
\begin{equation} \label{Equationsofmotion}
E^{\alpha}_{\beta}\doteq \dot{K}^{\alpha}_{\beta}-NKK^{\alpha}_{\beta}+
NR^{\alpha}_{\beta}+2N^{\xi}(C^{\alpha}_{\xi\rho}K^{\rho}_{\beta}
-C^{\rho}_{\xi\beta}K^{\alpha}_{\rho})-
K^{\alpha}_{\beta,m}\sigma^{m}_{\zeta}(x)N^{\zeta}
\end{equation}
\end{subequations}
and:
\begin{equation} \label{Kurvature}
K_{ij}=-\frac{1}{2N}\left(\dot{\gamma}_{\alpha\beta}+
2\gamma_{\alpha\nu}C^{\nu}_{\beta\rho}N^{\rho}+
2\gamma_{\beta\nu}C^{\nu}_{\alpha\rho}N^{\rho}\right)
\sigma^{\alpha}_{i}\sigma^{\beta}_{j}
\equiv K_{\alpha\beta}\sigma^{\alpha}_{i}\sigma^{\beta}_{j}
\end{equation}
\begin{equation} \label{Curvature}
\begin{split}
R_{ij}&=(C^{\kappa}_{\sigma\tau}C^{\lambda}_{\mu\nu}\gamma_{\alpha\kappa}\gamma_{\beta\lambda}
\gamma^{\sigma\nu}\gamma^{\tau\mu}+2C^{\lambda}_{\alpha\kappa}C^{\kappa}_{\beta\lambda}+
2C^{\mu}_{\alpha\kappa}C^{\nu}_{\beta\lambda}\gamma_{\mu\nu}\gamma^{\kappa\lambda}
+2C^{\lambda}_{\beta\kappa}C^{\mu}_{\mu\nu}\gamma_{\alpha\lambda}\gamma^{\kappa\nu}\\
&+2C^{\lambda}_{\alpha\kappa}C^{\mu}_{\mu\nu}\gamma_{\beta\lambda}\gamma^{\kappa\nu})
\sigma^{\alpha}_{i}\sigma^{\beta}_{j}+2C^{\epsilon}_{\mu\epsilon,j}\sigma^{\mu}_{i}
-C^{\epsilon}_{\mu\nu,l}\sigma^{l}_{\epsilon}\sigma^{\mu}_{i}\sigma^{\nu}_{j}\\
&-\sigma^{l}_{\epsilon_{1}}\gamma^{\epsilon_{1}\epsilon_{2}}\gamma_{\epsilon_{3}\epsilon_{4}}
C^{\epsilon_{3}}_{\epsilon_{5}\epsilon_{2},l}
\left(\sigma^{\epsilon_{4}}_{i}\sigma^{\epsilon_{5}}_{j}+\sigma^{\epsilon_{5}}_{i}\sigma^{\epsilon_{4}}_{j}\right)\equiv
R_{\alpha\beta}\sigma^{\alpha}_{i}\sigma^{\beta}_{j}
\end{split}
\end{equation}
as well as:
\begin{displaymath}
\begin{array}{cc}
K^{i}_{j}=g^{iw}K_{wj}=\sigma^{i}_{\alpha}\gamma^{\alpha\omega}
\sigma^{w}_{\omega}
K_{\rho\beta}\sigma^{\rho}_{w}\sigma^{\beta}_{j}=
\sigma^{i}_{\alpha}\sigma^{\beta}_{j}K^{\alpha}_{\beta}\\
\\
K=g^{ij}K_{ij}=\gamma^{\alpha\beta}K_{\alpha\beta}\\
\\
R=g^{ij}R_{ij}=C^{\alpha}_{\mu\kappa}C^{\beta}_{\nu\lambda}
\gamma_{\alpha\beta}\gamma^{\mu\nu}
\gamma^{\kappa\lambda}+2C^{\alpha}_{\beta\mu}C^{\beta}_{\alpha\nu}
\gamma^{\mu\nu}+
4C^{\alpha}_{\alpha\mu}C^{\beta}_{\beta\nu}\gamma^{\mu\nu}
-4C^{\alpha}_{\alpha\beta,m}\sigma^{\beta}_{n}g^{mn}
\end{array}
\end{displaymath}
It is note worthy that the quantities with Greek indices or no
indices, are time dependent only --although not manifestly.
Equation set (\ref{SpatialEinstein}), forms what is known as a
--complete-- perfect ideal; that is, there are no integrability
conditions obtained from this system. So, with the help of the
third of (\ref{Equationsofmotion}), (\ref{Kurvature}),
(\ref{Curvature}), it can explicitly be shown, that the time
derivatives of (\ref{Quadratic}) and (\ref{Linear}) vanish
identically. The calculation is straightforward --although
somewhat lengthy; the Jacobi identity
$C^{\alpha}_{\xi[\tau}C^{\xi}_{\rho\omega]}=0$ (of course, because
of the structure functions the appropriate Jacobi identity would,
in general, contain a term
$C^{\alpha}_{[\tau\rho,j}\sigma^{j}_{\omega]}$ which however
vanishes for structure functions here considered). The vanishing
of the time derivatives of the 4 constraint equations: $E_{0}=0,
E_{i}=0$, implies that these equations, are first integrals of
(\ref{Equationsofmotion}) --moreover, with vanishing integration
constants. Based on the intuition gained from the full theory, one
expects that the freedom which corresponds to these integrals, is
a reflection of the only known covariance of the theory; i.e. of
the freedom to make arbitrary changes of the time and space
coordinates.

Let us first consider the time reparameterization invariance;
if a transformation:
\begin{subequations}
\begin{equation}  \label{timereparametrization}
t\rightarrow \widetilde{t}=g(t) \Leftrightarrow t=f(\widetilde{t })
\end{equation}
is inserted in the line element (\ref{lineelementoneform}), it is
easily inferred that:
\begin{equation}
\gamma_{\alpha\beta}(t)\rightarrow
\gamma_{\alpha\beta}(f(\widetilde{t}))\equiv
\widetilde{\gamma}_{\alpha\beta}(\widetilde{t})
\end{equation}
\begin{equation}
\begin{split}
N(t)\rightarrow \pm~
N(f(\widetilde{t}))\frac{df(\widetilde{t})}{d\widetilde{t}}\equiv
\widetilde{N}(\widetilde{t})\\
N^{\alpha}(t)\rightarrow
N^{\alpha}(f(\widetilde{t}))\frac{df(\widetilde{t})}{d\widetilde{t}}\equiv
\widetilde{N}^{\alpha}(\widetilde{t})
\end{split}
\end{equation}
\end{subequations}
Accordingly, $K^{\alpha}_{\beta}$ transforms under
(\ref{timereparametrization}) as a scalar and thus (\ref{Quadratic}) and
(\ref{Linear}) are also scalar equations, while (\ref{Equationsofmotion}),
gets multiplied by a factor
$df(\widetilde{t})/d\widetilde{t}$. Thus, given a particular solution to
equations (\ref{SpatialEinstein}), one can always
obtain an equivalent solution, by arbitrarily redefining time. Hence, the existence of one arbitrary function of
time in the general solution to Einstein's equations (\ref{SpatialEinstein}),
is understood.

In order to understand the presence of the other arbitrary
functions of time, it is natural to turn our attention to the
transformations of the 3 spatial coordinates $x^{i}$. To begin
with, consider the transformations:
\begin{equation} \label{spatialtransformation}
\begin{split}
\widetilde{t}=t\Leftrightarrow& t=\widetilde{t}\\
\widetilde{x}^{i}=g^{i}(x^{j},t)\Leftrightarrow&
x^{i}=f^{i}(\widetilde{x}^{j},\widetilde{t})=f^{i}(\widetilde{x}^{j},t)
\end{split}
\end{equation}
Under these transformations, the line element (\ref{lineelementoneform})
becomes:
\begin{equation} \label{lineelementoneformtilded}
\begin{split}
ds^{2}&=\left(N^{\alpha}(t)N_{\alpha}(t)-N(t)^{2}+
2\sigma^{\alpha}_{i}(f)\frac{\partial f^{i}}{\partial
t}N_{\alpha}(t)+\frac{\partial f^{i}}{\partial t}\frac{\partial f^{j}}{\partial t
}~\sigma^{\alpha}_{i}(f)\sigma^{\beta}_{j}(f)
\gamma_{\alpha\beta}(t)\right)dt^{2}\\
&+2\sigma^{\alpha}_{i}(x)\frac{\partial f^{i}}{\partial
\widetilde{x}^{m}}\left(N_{\alpha}(t)+\sigma^{\beta}_{j}(x)\frac{\partial
f^{j}}{\partial t}\gamma_{\alpha\beta}(t)\right)d\widetilde{x}^{m}dt\\
&+\sigma^{\alpha}_{i}(x)\sigma^{\beta}_{j}(x)
\gamma_{\alpha\beta}(t)\frac{\partial
f^{i}}{\partial \widetilde{x}^{m}}\frac{\partial f^{j}}{\partial
\widetilde{x}^{n}}~d\widetilde{x}^{m}d\widetilde{x}^{n}
\end{split}
\end{equation}

Since our aim, is to retain spatial homogeneity of the line
element (\ref{lineelementoneform}), we have to refer the form of
the line element given in (\ref{lineelementoneformtilded}) to the old
basis $\sigma^{\alpha}_{i}(\widetilde{x})$ at the new spatial
point $\widetilde{x}^{i}$. Since $\sigma^{\alpha}_{i}$ --both at
$x^{i}$ and $\widetilde{x}^{i}$--, as well as, $\partial f^{i}/
\partial \widetilde{x}^{j}$, are invertible matrices, there
always exists a non-singular matrix $\Lambda
^{\alpha}_{\mu}(\widetilde{x},t)$ and a triplet
$P^{\alpha}(\widetilde{x},t)$, such that:
\begin{equation} \label{lambdapi}
\begin{split}
\sigma^{\alpha}_{i}(x)\frac{\partial f^{i}}{\partial
\widetilde{x}^{m}}=&\Lambda^{\alpha}_{\mu}(\widetilde{x},t)
\sigma^{\mu}_{m}(\widetilde{x})\Rightarrow
\frac{\partial f^{i}}{\partial
\widetilde{x}^{m}}=\Lambda^{\alpha}_{\beta}(\widetilde{x},t)
\sigma^{\beta}_{m}(\widetilde{x})\sigma^{i}_{\alpha}(x)\\
\sigma^{\alpha}_{i}(x)\frac{\partial f^{i}}{\partial t}=&
P^{\alpha}(\widetilde{x},t)\Rightarrow
\frac{\partial f^{i}}{\partial t}=P^{\alpha}(\widetilde{x},t)
\sigma^{i}_{\alpha}(x)
\end{split}
\end{equation}

The above relations, must be regarded as definitions, for the
matrix $\Lambda ^{\alpha}_{\mu}$ and the triplet $P^{\alpha}$.
With these identifications, the line element (\ref{lineelementoneformtilded}) assumes the
form:
\begin{equation}
\begin{split}
ds^{2}&=\left((N_{\alpha}(t)+P_{\alpha}(\widetilde{x},t))(N^{\alpha}(t)+
P^{\alpha}(\widetilde{x},t))-
N(t)^{2}\right)dt^{2}\\
&+2\left(N_{\alpha}(t)+P_{\alpha}(\widetilde{x},t)\right)
\Lambda^{\alpha}_{\beta}(\widetilde{x},t)
\sigma^{\beta}_{m}(\widetilde{x})dtd\widetilde{x}^{m}\\
&+\gamma_{\alpha\beta}(t)\Lambda^{\alpha}_{\mu}(\widetilde{x},t)
\Lambda^{\beta}_{\nu}(\widetilde{x},t)
\sigma^{\mu}_{m}(\widetilde{x})\sigma^{\nu}_{n}(\widetilde{x})
d\widetilde{x}^{m}d\widetilde{x}^{n}
\end{split}
\end{equation}

Consistency requirements, such as the non trivial action of the
transformation under discussion, change the character of
(\ref{lambdapi}), from definitions to a set of first-order highly
non-linear Partial Differential Equations (PDEs) for the
unknown functions $f^{i}$. The existence of local solutions to
these equations is guaranteed by Frobenius theorem
\cite{Warner} as long as the necessary and sufficient conditions:
\begin{displaymath}
\frac{\partial}{\partial \widetilde{x}^{j}}\Big( \frac{\partial
f^{i}}{\partial \widetilde{x}^{m}}\Big)- \frac{\partial}{\partial
\tilde{x}^{m}}\Big( \frac{\partial f^{i}}{\partial
\widetilde{x}^{j}}\Big)=0
\end{displaymath}
\begin{displaymath}
\frac{\partial}{\partial t}\Big( \frac{\partial
f^{i}}{\partial \widetilde{x}^{m}}\Big)- \frac{\partial}{\partial
\widetilde{x}^{m}}\Big( \frac{\partial f^{i}}{\partial
t}\Big)=0
\end{displaymath}
hold. These integrating conditions, with repeated use of
(\ref{DeRham}), result in:
\begin{equation} \label{systemlambda}
\begin{split}
\Lambda^{\xi}_{\beta,m}(\widetilde{x},t)\sigma^{\beta}_{n}(\widetilde{x})
-\Lambda^{\xi}_{\beta,n}(\widetilde{x},t)
\sigma^{\beta}_{m}(\widetilde{x})&=
2(C^{\xi}_{\alpha\beta}(x)\Lambda^{\alpha}_{\mu}(\widetilde{x},t)
\Lambda^{\beta}_{\nu}(\widetilde{x},t)\\
&-C^{\omega}_{\mu\nu}(\widetilde{x})
\Lambda^{\xi}_{\omega}(\widetilde{x},t))\sigma^{\mu}_{m}(\widetilde{x})
\sigma^{\nu}_{n}(\widetilde{x})
\end{split}
\end{equation}
\begin{equation} \label{systempi}
\frac{1}{2}P^{\alpha}_{,m}(\widetilde{x},t)=\left(\frac{1}{2}
\dot{\Lambda}^{\alpha}_{\beta}(\widetilde{x},t)-
C^{\alpha}_{\mu\nu}(x)P^{\mu}(\widetilde{x},t)
\Lambda^{\nu}_{\beta}(\widetilde{x},t)\right)
\sigma^{\beta}_{m}(\widetilde{x})
\end{equation}

The line element (\ref{lineelementoneformtilded}) can be written, more
concisely:
\begin{equation} \label{consiceLineelement}
ds^{2}\equiv (\widetilde{N}^{\alpha}\widetilde{N}_{\alpha}-
\widetilde{N}^{2})dt^{2}+2\widetilde{N}_{\alpha}\sigma^{\alpha}_{i}
(\widetilde{x})d\widetilde{x}^{i}dt
+\widetilde{\gamma}_{\alpha\beta}\sigma^{\alpha}_{i}
(\widetilde{x})\sigma^{\beta}_{j}(\widetilde{x})
d\widetilde{x}^{i}d\widetilde{x}^{j}
\end{equation}
with the allocations:
\begin{subequations} \label{actions}
\begin{equation} \label{allocationgamma}
\widetilde{\gamma}_{\alpha\beta}=\Lambda^{\mu}_{\alpha}(\widetilde{x},t)
\Lambda^{\nu}_{\beta}(\widetilde{x},t)
\gamma_{\mu\nu}(t)
\end{equation}
\begin{equation} \label{allocationpi}
\widetilde{N}_{\alpha}=\Lambda^{\beta}_{\alpha}(\widetilde{x},t)
(N_{\beta}(t)+P^{\rho}(\widetilde{x},t)
\gamma_{\rho\beta}(t))~\textrm{and thus}
~\widetilde{N}^{\alpha}=S^{\alpha}_{\beta}(\widetilde{x},t)
(N^{\beta}(t)+P^{\beta}(\widetilde{x},t))
\end{equation}
\begin{equation} \label{allocationlapse}
\widetilde{N}(t)=N(t)
\end{equation}
\end{subequations}
(where $S=\Lambda^{-1}$)

In order for the transformation (\ref{spatialtransformation}) to
preserve the splitting of the 3-metric, into spatial and temporal
parts, i.e.
$\widetilde{\gamma}_{\alpha\beta}=\widetilde{\gamma}_{\alpha\beta}
(t)$ in (\ref{allocationgamma}), one must restrict to the case
$\Lambda^{\alpha}_{\beta}(\widetilde{x},t)=\Lambda^{\alpha}_{\beta}(t)$. Indeed,
such a partial --because of space independence-- solution to
the system (\ref{systemlambda}), exists:
\begin{equation} \label{solutionlambda}
\Lambda^{\alpha}_{\beta}(t)=\left(
\begin{array}{ccc}
\lambda_{1}(t) & \lambda_{2}(t) & 0\\
0 & \lambda_{5}(t) & 0\\
0 & \lambda_{8}(t) & \lambda_{9}(t)
\end{array}
\right)
\end{equation}
under the condition:
\begin{equation} \label{restriction}
x^{2}=ArcCot(\frac{Cot(\widetilde{x}^{2})}{\lambda_{5}(t)})
\end{equation}
i.e.: $x^{2}=f^{2}(\widetilde{x}^{2},t)$, as far as the
transformation (\ref{spatialtransformation}) is concerned. This is
the automorphism group for the KS Type models. Then, equations
(\ref{systempi}) --in view of (\ref{restriction})-- result in:
\begin{subequations} \label{solution}
\begin{equation} \label{Lambdasolution}
\Lambda^{\alpha}_{\beta}(t)=\left(
\begin{array}{ccc}
\lambda_{1} & \lambda_{2}(t) & 0\\
0 & \lambda_{5} & 0\\
0 & \lambda_{8}(t) & \lambda_{9}
\end{array} \right)
\end{equation}
\begin{equation} \label{Pisolution}
P^{\alpha}(\widetilde{x},t)=\left(\lambda_{2}'(t)\widetilde{x}^{2}+h_{1}(t),
0,Sin(\widetilde{x}^{2})\left(h_{2}(t)+\lambda_{8}'(t)
Ln\left(Tan(\frac{\widetilde{x}^{2}}{2})\right)\right)\right)
\end{equation}
\end{subequations}
where the prime, denotes temporal differentiation and
$\lambda_{2}(t)$, $\lambda_{8}(t)$, $h_{1}(t)$, $h_{2}(t)$, are
arbitrary functions of time.

At this point, it must be observed that spatial homogeneity implies that
the Ricci scalar is space independent, on a given --albeit arbitrary--
hypersurface of simultaneity $\Sigma_{t}$.
Through computing facility, one may see that this requirement, restricts
the possible admissible forms of the scale factor
matrix $\gamma_{\alpha\beta}(t)$, to the set:
\begin{equation} \label{admissiblemetric}
\gamma_{\alpha\beta}(t)=\left(
\begin{array}{ccc}
\gamma_{11}(t) & \gamma_{12}(t) & 0\\
\gamma_{12}(t) & \gamma_{22}(t) & \gamma_{23}(t)\\
0 & \gamma_{23}(t) & \gamma_{33}(t)
\end{array}
\right)
\end{equation}

We now conclude this section, by describing the --somehow lengthly but mostly
algebraic-- algorithm of the application
of these results to the most general admissible 3-metric for the KS vacuum.\\
To this end, consider the most general metric of type (\ref{admissiblemetric}),
with a lapse function $N(t)$ and a full shift vector
$N^{\alpha}(t)$ present, in a system of local coordinates
$(t,r,\theta,\phi)$, and a transformation of the type
(\ref{spatialtransformation}) --under the condition
(\ref{restriction})-- to a new set of local coordinates
$(t,\widetilde{r},\widetilde{\theta},\widetilde{\phi})$. In this
new ``frame'', the lapse function, the shift vector and the
3-metric, can be found using the transformation laws
(\ref{allocationgamma}), (\ref{allocationpi}), (\ref{allocationlapse}), where
$\Lambda^{\alpha}_{\beta}(t)$ and $P^{\alpha}(\widetilde{x},t)$ are to be
calculated from (\ref{solution}).\\
Using the freedom provided by the temporal functions $\lambda_{2}(t)$ and
$\lambda_{8}(t)$, the initial scale factor matrix is brought to a diagonal
form, called ``final scale factor matrix'', --which depends on some
combinations of the elements of the initial scale factor matrix.
The corresponding alterations to the lapse function and the shift vector,
can also be found; the lapse function remains unchanged and the ``final shift
vector'', depends on some combinations of the initial shift vector, and the
elements of the initial scale factor matrix.\\
Inserting these ``final quantities'' into the linear constraints
(\ref{Linear}), it is found that the shift vector must be zero --expect for
its first component, in which the EFEs are ``transparent'' and thus at one's
disposal-- and some correlations on the combinations of the
elements of the initial scale factor matrix appearing, which bring the diagonal
``final scale factor matric'', to the form
$\textrm{diag}(\gamma_{11}(t),\gamma_{22}(t),\gamma_{22}(t))$.
Our purpose, is thus, fulfilled. Starting from the most general, admissible
form of the 3-metric and a shift vector, through a particular class of GCTs,
we end up to the KS metric, with no shift. This fact, is an alternative proof
of not only the generality, but also of the irreducibility of the KS vacuum
solution.

Our method makes it possible to give the KS vacuum metric in the most general
form containing all the relevant ``gauge'' freedom; one has only to solve
the reduced system and invert the transformation which enables the reduction.
The solution of the reduced system is (in the time gauge $\gamma_{22}(t)=t^{2}$):
\begin{equation} \label{KSvacuum}
ds^{2}=-(\frac{C}{t}-1)^{-1}dt^{2}+(\frac{C}{t}-1)dr^{2}+t^{2}d\theta^{2}
+t^{2}Sin(\theta)^{2}d\phi^{2}
\end{equation}
where $C$, is a constant.
Therefore, using (\ref{solution}) one can write the most general metric --in
matrix notation-- of the KS type:
\begin{equation} \label{Closed1}
\gamma_{\textrm{Most General}}(t)=\Lambda^{T}(t)\gamma_{KS}(f(t))\Lambda(t)
\end{equation}
where $\Lambda^{\alpha}_{\beta}$ is to be given by (\ref{Lambdasolution}) and:
\begin{equation} \label{Closed2}
N(t)=f'(t)(\frac{C}{f(t)}-1)^{-1/2}
\end{equation}
\begin{equation} \label{Closed3}
N^{\alpha}(t)=S^{\alpha}_{\beta}(t)P^{\beta}(f(t))
\end{equation}
where $P^{\beta}$ is to be given by (\ref{Pisolution}) and $S=\Lambda^{-1}$.
\section{Discussion}
We have considered the set of GCTs which preserve manifest spatial
homogeneity for the case of symmetry groups acting multiply
transitively on the hypersurfaces of simultaneity $\Sigma_{t}$
(\ref{Killing}), (\ref{LieAlgebra}). This implies that the
associated basis 1-forms satisfy an open algebra (\ref{DeRham}).
The transformations found (\ref{spatialtransformation}),
(\ref{lambdapi}) mix space and time in the new space variables,
their existence is guaranteed by the Frobenious Theorem
(\ref{systemlambda}), (\ref{systempi}) and their effect on the
general spatially homogeneous line element
(\ref{lineelementoneform}) and/or (\ref{consiceLineelement}),
(\ref{actions}) can be used to simplify the reduced EFEs,
(\ref{SpatialEinstein}). In this way an alternative proof is
given, of the fact that the well known KS vacuum metric
(\ref{KSvacuum}) is the irreducible form of the most general
geometry admitting $G_{4}$ invariant homogeneous hypersurfaces.
The closed form of this metric (\ref{Closed1}), (\ref{Closed2}),
(\ref{Closed3}), exhibiting all the relevant ``gauge'' freedom in
arbitrary functions of time, is also given.

The merits of this analysis can be most easily brought forward by
comparing it to the original derivation: the 3-dimensional KVFs
(\ref{Killing}) are first trivially prolonged to spacetime vectors
by adding to each of them a vanishing time component; then one
demands for these vectors to be symmetries of the spacetime metric
sought for. This results in the well known form of the KS vacuum
metric (modulo a $g_{01}$ component). Implicit in the above
derivation is the following important assumption: the space
variables in which the original KVFs are given, are to be
identified to the spatial coordinates of the Gauss
normal system of the requested metric (an assumption which justifies the choice $g_{01}=0$, a posteriori).\\
On the contrary, the analysis here presented makes no such
assumption. The coordinates in which the original KVFs --and thus
the basis 1-forms-- are given can be any of the set obtained in
(\ref{spatialtransformation}). Consequently the scale factor
matrix can be filled with non vanishing elements --along with
the existence of non vanishing shift. Then the linear constraint
EFEs ensure that when, through the effect of a particular AID
(\ref{Lambdasolution}), the scale factor matrix is cast into a diagonal form,
the shift vanishes.

Lastly we would like to point out that our analysis is susceptible
to generalization for the cases of lesser symmetry, say when there
exist only two KVFs: one must first find a basis 1-forms
--invariant under the action of these KVFs--, assume a splitting
of the spatial metric of the form
$g_{ij}(t,x)=\gamma_{\alpha\beta}\sigma^{\alpha}_{i}(x)\sigma^{\beta}_{j}(x)$,
where the ``scale factor'' matrix will be supposed to depend not
only on $t$, but also on the unique combination of the $x^{i}$
that has zero Lie derivative with respect to the KVFs. Then the
integrability conditions will give spatial coordinate
transformations which hopefully will simplify the EFEs.
\vspace{1.5cm}

\textbf{\large{Acknowledgements}}\\
One of us (G.O. Papadopoulos) is currently a scholar of the Greek
State Scholarships Foundation (I.K.Y.) and acknowledges the
relevant financial support.

\end{document}